\documentclass[a4paper,fleqn,usenatbib]{mnras}
\usepackage{graphicx}
\usepackage{newtxtext,newtxmath}
\usepackage{url}
\usepackage[T1]{fontenc}
\usepackage{aecompl}

\newcommand{\zab}{\ensuremath{z_\textrm{\scriptsize abs}}}

\newcommand{\stat}{\ensuremath{\textrm{\scriptsize{stat}}}}
\newcommand{\syst}{\ensuremath{\textrm{\scriptsize{syst}}}}
\newcommand{\lab}{\ensuremath{\textrm{\scriptsize{lab}}}}
\newcommand{\obs}{\ensuremath{\textrm{\scriptsize{obs}}}}
\newcommand{\dmm}{\Delta\mu/\mu}
\newcommand{\kms}{km\,s$^{-1}$}
\newcommand{\ms}{m\,s$^{-1}$}
\newcommand{\jo}{J0000+0048}
\newcommand{\hi}{H\,\textsc{i}}
\newcommand{\civ}{C\,\textsc{iv}}
\newcommand{\htwo}{H$_{2}$}
\newcommand{\nco}{N_{\textrm{\scriptsize CO}}}
\newcommand{\bco}{b_{\textrm{\scriptsize CO}}}
\newcommand{\tco}{T_{\textrm{\scriptsize CO}}}
\newcommand{\astate}{\textrm{A}^{1} \Pi}
\newcommand{\xstate}{\textrm{X}^{1}\Sigma^{+}}

\def\revmodphys{Rev.\ Mod.\ Phys.}
\def\jqsrt{J.\ Quant.\ Spectrosc.\ Radiat.\ Transfer}

\title[Constraint on a varying $\mu$ at $\zab \simeq 2.5$ from CO]{Analysis of carbon monoxide absorption at $\zab \simeq 2.5$ to constrain variation of the proton-to-electron mass ratio}

\author[Dapr\`a et al.]{M. Dapr\`a,$^{1}$ P. Noterdaeme,$^{2}$ M. Vonk,$^{1}$ M. T. Murphy,$^{3}$ and W. Ubachs,$^{1}$\\
$^1$Department of Physics and Astronomy, LaserLaB, VU University,
  De Boelelaan 1081, 1081 HV Amsterdam, The Netherlands\\
$^{2}$Institut d'Astrophysique de Paris, CNRS-UPMC, UMR7095, 98bis bd Arago, 75014 Paris, France\\
$^{3}$Centre for Astrophysics and Supercomputing, Swinburne University of Technology, Melbourne, Victoria 3122, Australia}

\pubyear{2016}

\begin{document}
\label{firstpage}
\pagerange{\pageref{firstpage}--\pageref{lastpage}} 
\maketitle

\begin{abstract}
Absorption by carbon monoxide in the spectrum of quasar SDSS J000015.16+004833.2 is investigated in order to derive a constraint on the temporal variation of the proton-to-electron mass ratio, $\mu$. The spectrum was recorded using VLT/UVES, and it was partially corrected for long-range wavelength scale distortions using the supercalibration technique. Eight vibrational CO singlet-singlet bands belonging to the \mbox{$\astate$-$\xstate$} electronic absorption system, and the perturbing d$^{3}\Delta$-$\xstate(5, 0)$ singlet-triplet band are detected in the damped Lyman-$\alpha$ system at $\zab \simeq 2.52$. The spectra are modelled using a comprehensive fitting technique, resulting in a final value of $\dmm = (1.8 \pm 2.2_{\stat} \pm 0.4_{\syst}) \times 10^{-5}$, which is consistent with no variation over a look-back time of \mbox{$\sim 11.2$ Gyrs}.
\end{abstract}


\section{Introduction}
\label{sec:intro}
The search for a temporal variation proton-to-electron mass ratio, $\mu \equiv M_{P}/m_{e}$, is performed by investigating molecular absorption in high-redshift systems. \cite{Thompson1975} first suggested to probe a possible variation of $\mu$ using the absorption of molecular hydrogen, \htwo, and carbon monoxide, CO, in quasar spectra. More recently, a wide variety of molecules was found to be sensitive to a variation of $\mu$ \citep{Jansen2011,Jansen2011b,Jansen2014}. Observations of ammonia \citep{Murphy2008,Kanekar2011} and methanol \citep{Bagdonaite2013a,Bagdonaite2013b,Kanekar2015} molecules in the radio domain returned a constraint \mbox{$|\dmm| < 10^{-7}$ ($1 \sigma$)} from two absorption systems at \mbox{$\zab < 1$}. Molecular hydrogen is a target for constraining $\mu$-variation in absorbing systems at redshifts \mbox{$\zab > 2$}. The analysis of \htwo\ absorption in the ten best absorbers, in terms of brightness (Bessel $R_{mag} \le 18.4$) and \htwo\ column density ($\log[N / \textrm{cm}^{-2}] \ge 14.5$), delivered the constraint of \mbox{$|\dmm| < 5 \times 10^{-6}$ ($3 \sigma$)} covering a window of look-back times of \mbox{$\sim 10.5$-12.5 Gyrs} \citep{Ubachs2016}.

Carbon monoxide, the second most abundant molecule in the Universe, is another target for a $\mu$-variation analysis. In particular, its $\astate-\xstate$ band system was detected in 8 absorbers at absorption redshifts $\zab > 1$: SDSS J160457.50+220300.5 \citep{Noterdaeme2009}, SDSS J085726.78+185524.3, SDSS J104705.75+205734.5, SDSS J170542.91+354340.2 \citep{Noterdaeme2011}, SDSS J143912.04+111740.5 \citep{Srianand2008},  SDSS J121143.42+083349.7 \citep{Ma2015}, SDSS J123714.60+064759.5 \citep{Noterdaeme2010}, and SDSS J000015.16+004833.2, hereafter \jo, \citep{Noterdaeme2017}. \cite{Salumbides2012} reported high-accuracy laboratory wavelength measurements, with an accuracy of $\Delta\lambda/\lambda = 3 \times 10^{-7}$, for this band system, allowing to extract  a constraint on a varying $\mu$ at a level of $\sim 10^{-5}$. \cite{Dapra2016} reported the first measurement of \mbox{$\dmm = (0.7 \pm 1.6_{\stat} \pm 0.5_{\syst}) \times 10^{-5}$} from electronic CO absorption only in the system SDSS~J123714.60+064759.5, which was then combined with the $\dmm$ value derived from \htwo\ absorption.

\cite{Noterdaeme2017} presented a detailed study of the absorption system towards \jo, including CO, H$_2$, deuterated molecular hydrogen, HD, as well as atomic species and dust, in order to understand the chemical and physical properties of the absorbing gas. Here, a new analysis of the CO lines in this system is presented, specifically focused on deriving a constraint on the temporal variation of $\dmm$ over cosmological timescales. The observations used in this work are presented in \mbox{Section \ref{sec:data}}, the absorption model and the derived value of $\dmm$ are presented in \mbox{Section \ref{sec:analysis}}, and the discussion of the systematic uncertainty is given in \mbox{Section \ref{sec:systematics}}.

\section{Data}
\label{sec:data}
Quasar \jo\ was observed using the Ultraviolet and Visual Echelle Spectrograph \citep[UVES, ][]{Dekker2000} mounted on the 8.2 m Very Large Telescope (VLT) in two programmes: 093.A-0126(A) in 2014 (PI Paris), and 096.A-0354(A) in 2015 (PI Noterdaeme). The former programme was carried out using the 390+564 dichroic setting and a slit width of \mbox{0.9 arcsec} (resolving power $R \sim 50\,000$), while the latter used the same settings but a narrower slit width of \mbox{0.7 arcsec} ($R \sim 65\,000$) in the red arm of UVES. The work presented here is based on the full set of UVES exposures presented by \cite{Noterdaeme2017} and summarized in \mbox{Table \ref{tab:exposures}}.
\begin{table*}
 \centering
    \caption{Observational details of the \jo\ exposures with UVES/VLT used in this work.}
    \label{tab:exposures}
    \begin{tabular}{ccccccc}
    \hline
    Programme ID & Date & Exposure & Grating & \multicolumn{2}{c}{Slit width [arcsec]} & Dedicated \\
     &  & time [s] & [nm] & Blue arm & Red arm & supercalibration \\
    \hline
    \hline
    093.A-0126(A) & 02-08-2014 & 4800 & 390+564 & 0.9 & 0.9 & No \\
    093.A-0126(A) & 03-08-2014 & 4800 & 390+564 & 0.9 & 0.9 & No \\
    093.A-0126(A) & 04-08-2014 & 1934 & 390+564 & 0.9 & 0.9 & No \\
    093.A-0126(A) & 05-08-2014 & 4800 & 390+564 & 0.9 & 0.9 & No \\
    093.A-0126(A) & 20-08-2014 & 4800 & 390+564 & 0.9 & 0.9 & No \\
    093.A-0126(A) & 21-08-2014 & 4800 & 390+564 & 0.9 & 0.9 & No \\
    \hline
    096.A-0354(A) & 11-10-2015 & 4200 & 390+564 & 0.9 & 0.7 & Yes \\
    096.A-0354(A) & 03-11-2015 & 4200 & 390+564 & 0.9 & 0.7 & Yes \\
    096.A-0354(A) & 09-11-2015 & 4200 & 390+564 & 0.9 & 0.7 & Yes \\
    096.A-0354(A) & 12-11-2015 & 4200 & 390+564 & 0.9 & 0.7 & Yes \\
    096.A-0354(A) & 13-11-2015 & 4200 & 390+564 & 0.9 & 0.7 & Yes \\
    096.A-0354(A) & 13-11-2015 & 4200 & 390+564 & 0.9 & 0.7 & Yes \\
    096.A-0354(A) & 01-12-2015 & 4200 & 390+564 & 0.9 & 0.7 & Yes \\
    \hline
    096.A-0354(A) & 09-11-2015 & 4800 & 437+760 & 0.9 & 0.9 & No \\
    096.A-0354(A) & 09-11-2015 & 4800 & 437+760 & 0.9 & 0.9 & No \\
    \hline
    \end{tabular}
\end{table*}

The raw 2D exposures were reduced following the same procedure as in \cite{Bagdonaite2014}. The Common Pipeline Language (CPL) version of the UVES pipeline was used first to flat-field and bias-correct the exposures, and then to optimally extract the quasar flux. Each quasar exposure was wavelength calibrated using the standard `attached' ThAr exposure taken immediately after the science exposure. In addition to the standard wavelength calibration, the exposures recorded in 2015 were followed by a `supercalibration' exposure of the solar twin star HD001835 taken with the same grating settings as for quasar observation (see \mbox{Section \ref{subsubsec:longrange}}). After the standard reduction, the custom software \textsc{UVES\_popler} \citep{UVESpopler} was used to combine the echelle orders onto a common vacuum-heliocentric wavelength grid. The spectral resolution is \mbox{$\sim 6$ \kms} and \mbox{$\sim 4.5$ \kms} for exposures taken in 2014 and in 2015, respectively. In order not to undersample the latter exposures, a dispersion of \mbox{2.0 \kms} per pixel was used for the wavelength grid. This software was also used to identify and remove `bad' pixels and other spectral artifacts and to fit a continuum using low-order polynomials.

The final \jo\ spectrum, after the reduction, covers the wavelengths from 3284.2 to \mbox{9465.8 \AA}, with gaps between 5615.6-5671.1 and \mbox{7525.7-7655.7 \AA} due to the CCD separations. The signal-to-noise ratio (S/N) is $\sim$ 19 per \mbox{2.0 \kms} per pixel at \mbox{$\sim 5000$ \AA}, in the middle of the CO window in the spectrum.

\section{Carbon monoxide absorption}
\label{sec:analysis}
CO absorption in the spectrum of \jo\ was first reported by \cite{Noterdaeme2017}, who detected 9 CO bands belonging to the $\astate(\nu\,'=0$-8) - $\xstate(\nu\,''=0)$ band system and one to the d$^{3}\Delta(\nu\,'=5$) - $\xstate(\nu\,''=0)$ inter-band system. They found that the A-X(5-0) band is completely blended with Si~\textsc{iv} absorption at $\zab = 2.52$, while the A-X(2-0) band is overlapped in its R and Q branches by C~\textsc{iv} absorption at $\zab = 2.36$. Similarly, the A-X($\nu\,'=6$, 7-0) bands have their P branches overlapped by intervening \hi\ lines from the Lyman-$\alpha$ forest.

\subsection{Fitting method}
\label{subsec:method}
The approach developed by \cite{Dapra2016} was used to model the CO absorption in the quasar spectrum. A vibrational contour for each CO band was computed and fitted to the spectrum using the comprehensive fitting technique, introduced by \cite{King2008} and later refined by \cite{Malec2010}. This method involves a simultaneous treatment of all the transitions, achieved by tying some of the fitting parameters together, which results in a lower number of free parameters in the fit. Moreover, the comprehensive fitting technique allows to handle the overlaps of CO absorption features with intervening spectral features, like \hi\ lines from the Lyman-$\alpha$ forest and narrow metal lines, as well as the blending of the different branches of the CO bands.

To create the absorption model, the non-linear least-squares Voigt profile fitting program \textsc{vpfit} \citep{vpfit} was used. In \textsc{vpfit} a Voigt profile is described by a set of free parameters describing the properties of the absorbing system, and a set of fixed values describing the atomic and molecular properties. The free parameters are the column density \emph{N}, the absorption redshift $\zab$, and the Doppler line-width \emph{b}. The vibrational contours for the CO bands were created by tying together the free parameters for each CO transition. The fixed values are the laboratory wavelength $\lambda^{0}$, the oscillator strength \emph{f}, the damping parameter $\Gamma$, and the sensitivity coefficient \emph{K}. For the CO molecule, these parameters are summarised in the database reported by \cite{Dapra2016}.

\subsection{Temperature determination}
\label{subsec:temperature}
The band contours were created, assuming thermodynamic equilibrium, by linking the column densities $N_{J}$ using a temperature-dependent partition function given by:
\begin{equation}
    N_{J}(T) = \nco P_{J}(T) = \nco \frac{(2J+1)e^{-E_{\textrm{\scriptsize{rot}}}/k\tco}}{\sum(2J+1)},
    \label{eq:temperature}
\end{equation}
where $\nco$ is the total CO column density, and $P_{J}(T)$ is the partition function determining the population distribution over the rotational levels.

Since \textsc{vpfit} does not include the temperature as a fitting parameter, it was determined by fitting multiple absorption models, corresponding to different $\tco$ values. At this stage of the analysis, only the non-overlapping CO bands were considered. This was done to exclude from the absorption model any non-CO feature, in order to avoid that the $\tco$ value was affected by uncertainties in the modelling of such features.

\begin{figure}
    \centering
    \includegraphics[width=\columnwidth]{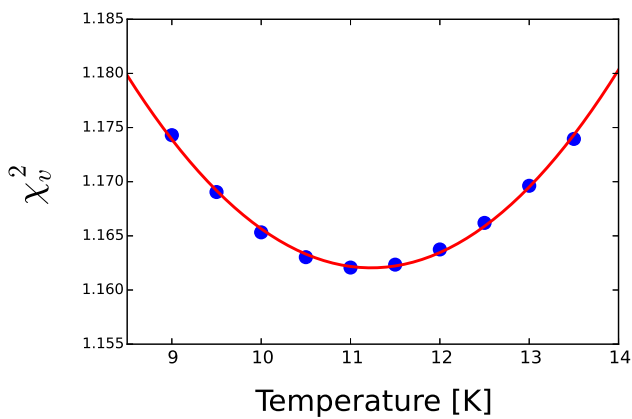}
    \caption{Reduced $\chi^{2}$ values returned by the CO models with different temperatures. The (blue) dots show the $\chi^{2}_{\nu}$ values and the (red) solid line indicates the best fit.}
    \label{fig:temperature}
\end{figure}

The reduced chi squared values, $\chi^{2}_{\nu}$, as a function of the temperature yielded, as shown in \mbox{Fig. \ref{fig:temperature}}, a statistically preferred value of \mbox{$\tco = 11.4  \pm 0.4$ K}. This represents an excitation temperature averaged over $J = 0$ to $J = 5$ levels. This is not the same temperature as measured by \cite{Noterdaeme2017}, who did not assume thermodynamic equilibrium, using only the excitation of the lower rotational levels, i.e. $J=0$ to $J=3$, to derive the CMB temperature, after correction for collisional excitation. Indeed, they also showed that higher-$J$ levels, which exhibit much larger energy differences, appear to slightly deviate from thermodynamical equilibrium. 
Here, no attempt has been made to derive a true physical temperature in connection to $T_{\textrm{\scriptsize{CMB}}}$ and including collisional excitation effects, but rather to determine an optimised representation of the band contours with the aim of an accurate wavelength measurement for the CO absorption. The latter is crucial for extracting information on varying constants. It is also noted that the uncertainty on $\tco$ is purely statistical and does not cover systematic effects due to the neglect of collisional excitation effects and the assumption of thermodynamic equilibrium. The systematic consequences of the assumptions on the final value of $\dmm$ are discussed in \mbox{Section \ref{subsec:temp_err}}.

\subsection{Absorption model}
\label{subsec:model}
After a value of $\tco$ was determined, the population distribution was fixed via \mbox{Eq. (\ref{eq:temperature})}. This means that thermodynamic equilibrium is imposed in this analysis for fitting the absorption model against the quasar spectrum. Since the comprehensive fitting technique can handle overlaps among absorption features, the CO bands that are partially overlapped by intervening lines were included in the model. This does not hold for the A-X($5-0$) band, which is completely overlapped by metal absorption. Since no relevant information is gained in such a case of complete overlap, this CO band was not considered in this work. Finally, it is noted that the A-X($8-0$) band lies towards the bluer part of the spectrum, where the absolute flux is lower. This is reflected by the lower S/N of $\sim 10$ per \mbox{2.0 \kms} per pixel at \mbox{$\sim 4700$} \AA.

The intervening \hi\ lines that are partially overlapping the \mbox{A-X($\nu\,' = 6$, $7-0$)} bands were included in the absorption model. A set of free parameters, \emph{N}, $\zab$, and \emph{b}, was assigned to each of these lines in \textsc{vpfit}. Since no assumptions were made about the origin of these \hi\ absorption features, their corresponding parameters were left untied and free to vary independently from each other. The A-X($2-0$) band is partially overlapped by the shorter wavelength component of the \civ\ doublet (laboratory wavelength: \mbox{$\lambda_{\lab} = 1548.20$ \AA}) absorbing at the redshift $\zab \simeq 2.36$. To handle the overlap and properly model the \civ, the longer wavelength component (\mbox{$\lambda_{\lab} = 1548.20$ \AA}) of the doublet was included in the absorption model. \civ\ shows a complex absorption profile, featuring multiple velocity components (VCs) that were modelled by assigning to each of them a set of free parameters. Each VC originates at a slightly different $\zab$ and, in principle, is observed under different physical conditions. To accommodate this, the fitting parameters of different VCs were not tied together. 

\cite{Noterdaeme2017} investigated the effect of continuum placement uncertainties by `shaking' the continuum level, finding that $\tco$ strongly depends on the continuum placement. However, shifts of the continuum affect only the fitting parameters $\nco$ and \emph{b}, having little effect on $\dmm$. To account for possible quasar continuum misplacements, a continuum correction was included in each spectral region considered. Such correction locally applies a constant rescaling of the global continuum, therefore minimising the impact of any global misplacement. Since the CO A-X($4-0$) band falls on top of the quasar N~\textsc{v} emission line, a continuum correction including an extra linear term beside the constant rescaling term was applied to the spectral region containing this band. 

\begin{figure}
    \centering
    \includegraphics[width=\columnwidth]{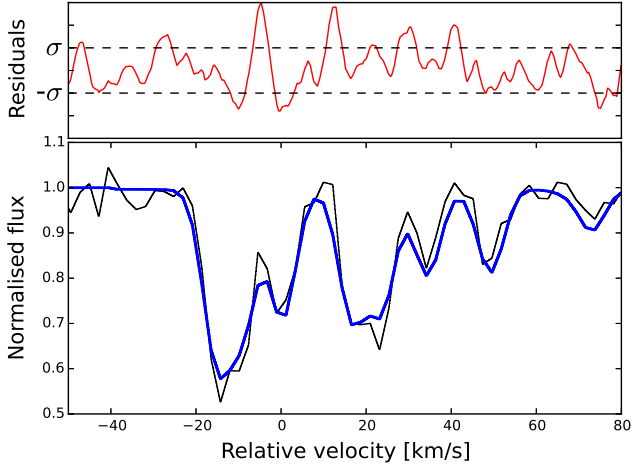}
    \caption{Top panel: normalized CRS from 6 non-overlapping CO bands. The dashed lines represent the $\pm 1 \sigma$ boundaries, which correspond to $\pm 0.06$ on the scale of the bottom panel. Bottom panel: the CO A-X(4-0) band is plotted as reference. The velocity scale is centred at the absorption redshift $\zab = 2.525464$.}
    \label{fig:crs}
\end{figure}

The $\chi^{2}_{\nu}$ parameter returned by the best-fit model is \mbox{$\chi^{2}_{\nu} = 1.1$}, which is slightly larger than unity. This may be due to the presence of extra, unresolved CO VCs that were not included in the model. The presence of such VCs was investigated using a composite residual spectrum \citep[CRS,][]{Malec2010} built by combining the residuals of the 6 non-overlapping CO bands. The CRS, which is presented in \mbox{Fig. \ref{fig:crs}}, shows possible evidence for extra VCs in the CO absorption profile (specifically, the \mbox{$> 1.5 \sigma$} deviations at velocities $\sim -5$ and \mbox{$\sim 15$ \kms}). Multiple 2 VCs models were fitted, resulting either in the rejection of the extra VC in \textsc{vpfit} or in significantly higher $\chi^{2}_{\nu}$ parameters. Thus, the presence of a second VC was excluded from the absorption model.

\begin{figure*}
    \centering
    \includegraphics[width=2\columnwidth]{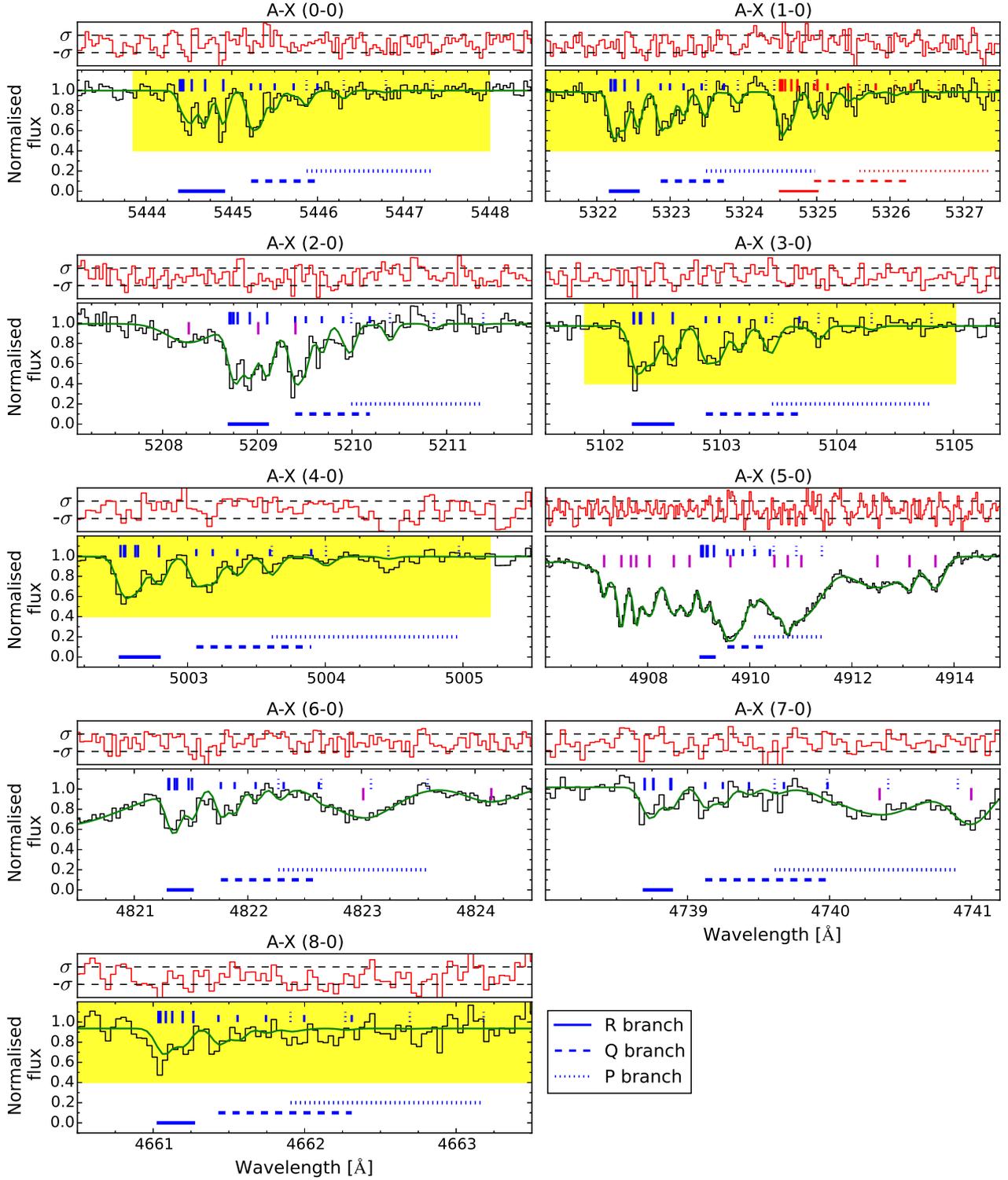}
    \caption{Absorption model for the CO bands considered in this work. The (green) solid line shows the fitted model, while the (blue) ticks indicate the wavelengths of the rotational lines for ground states $J=0-5$ and their different branches. The residuals, as well as their \mbox{$\pm 1 \sigma$} boundaries, are shown by the (red) solid line plotted above each panel. The value of $\sigma$ in each upper panel corresponds numerically to $\sim 0.06$ in terms of normalised flux. Band A-X(1-0) is perturbed by the inter-system band d-X(5-0), indicated by the (red) ticks. The extent of the R, Q and P branches is shown by the horizontal solid, dashed, and dotted lines respectively. The intervening absorption features that are overlapping the CO bands are indicated by solid (magenta) ticks. The band A-X($2 - 0$) is overlapped by C\,\textsc{iv} absorption features at $\zab \simeq 2.36$, band A-X($5 - 0$) is overlapped by Si\,\textsc{iv} absorption features ($\zab \simeq 2.52$), and bands with $\nu'=6$ and 7 are overlapped by \hi\ lines from the Lyman-$\alpha$ forest. The (yellow) shaded area shows the spectral regions used to derive the value of $\tco$.}
    \label{fig:bands}
\end{figure*}

The best-fit model, shown in \mbox{Fig. \ref{fig:bands}}, returned a total CO column density of $\log[\nco/\textrm{cm}^{-2}] = 15.00 \pm 0.04$, an absorption redshift of $\zab = 2.525464 \pm 0.000003$, and a line width \mbox{$b = 0.84 \pm 0.04$ \kms}, where the quoted errors represent the $1 \sigma$ uncertainties. 
%
Excluding the weak A-X($8-0$) band from the model does not affect the statistical uncertainties returned by the fit. This is considered evidence of the robustness of the comprehensive fitting technique against the larger residuals of the A-X($8-0$) band. 
%
The total column density is in good agreement with that reported by \cite{Noterdaeme2017}, while the Doppler width is \mbox{$\sim 18\%$} larger than the value reported by \cite{Noterdaeme2017}. The latter difference is due to the deviation from thermodynamic equilibrium of the high-\emph{J} rotational levels, which are not corrected for collisional excitation. As for $\tco$, the Doppler parameter \emph{b} is ascribed to the turbulent motions in the absorber rather than representing a kinetic temperature.

\subsection{Constraining $\dmm$}
\label{subsec:dmm}
The rotational states of the detected CO bands are sensitive to a variation of $\mu$, which will cause a shift of the wavelengths at which such lines are detected. This shift, which is assumed to have a linear dependence on  a varying $\mu$, is given by:
\begin{equation}
    \lambda^{\obs}_{i} = \lambda^{\lab}_{i}(1+\zab)(1+K_{i}\frac{\Delta\mu}{\mu}),
    \label{eq:shift}
\end{equation}
where $\lambda^{\obs}_{i}$ is the observed wavelength of the \emph{i}-th transition, $\lambda^{\lab}_{i}$ its rest wavelength, $\zab$ is the redshift at which absorption occurs, $\dmm \equiv (\mu_{\textrm{\scriptsize{z}}}-\mu_{\lab})/\mu_{\lab}$ is the relative difference between the proton-to-electron mass ratio measured in the absorbing system, $\mu_{\textrm{\scriptsize{z}}}$, and in the laboratory, $\mu_{\lab}$, and $K_{i}$ is the sensitivity coefficient of the \emph{i}-th transition. The sensitivity coefficients express the sign and magnitude of the sensitivity to a varying $\mu$ and are specific for each transition.

The $\dmm$ value was calculated in \textsc{vpfit} by adding an extra free parameter to the set describing the CO absorption. This fourth free parameter was added only after a robust absorption model was developed in order to avoid that an artificial $\mu$-variation compensated a flaw in the model itself. The model returned a value of $\dmm =  (2.2 \pm 2.2_{\stat}) \times 10^{-5}$, hereafter referred to as the fiducial value. The statistical error is derived only from the diagonal terms of the final covariance matrix for the fit, and it represents the statistical uncertainty ($1 \sigma$) in $\dmm$ derived from the S/N of the quasar spectrum.

\section{Systematic uncertainty}
\label{sec:systematics}

\subsection{Temperature choice}
\label{subsec:temp_err}
The fiducial value was derived assuming a Boltzmann level population determined by a CO temperature \mbox{$\tco = 11.4$ K}. The uncertainty on the fiducial value of $\dmm$ introduced by the temperature choice was tested by imposing \mbox{$\tco = 9.9$ K}, as reported by \cite{Noterdaeme2017}. The $\dmm$ value returned by the model built starting from this CO temperature is \mbox{$\dmm = (2.3 \pm 2.2_{\stat}) \times 10^{-5}$}. The difference of \mbox{$\sim 0.1 \times 10^{-5}$} between this value and the fiducial one was interpreted as the contribution to the total systematic uncertainty due to the temperature determination, and was added to the systematic error budget.

\subsection{Velocity shift between UVES arms}
\label{subsec:vel_shift}
The CO bands used in this work fall in the red arm in all exposures taken with grating settings \mbox{390+564 nm}, while they are partially covered by the blue arm in the two exposures taken with the \mbox{437+760 nm} grating settings (see \mbox{Table \ref{tab:dist}} for grating settings during exposures). The presence of a velocity offset between the blue and the red arm of UVES could introduce a systematic error on the fiducial value of $\dmm$ when combining the exposures.

To estimate the impact of such a shift, a sub-spectrum was built using only the exposures recorded in setting \mbox{390+564 nm}. The same procedures described in \mbox{Section \ref{sec:data}} for combining the exposures were followed, but the \mbox{437+760 nm} exposures were then removed and the spectrum recombined. A value of $\dmm = (2.4 \pm 2.2_{\stat}) \times 10^{-5}$ was extracted from the sub-spectrum and its deviation from the fiducial value was interpreted as the effect of a velocity shift between the arms of UVES. Therefore, an uncertainty on $\dmm$ of \mbox{$\sim 0.2 \times 10^{-5}$} was added to the systematic error budget.

\subsection{Wavelength scale distortions}
\label{subsec:distortions}
An accurate wavelength calibration of the quasar exposures is crucial in order to constrain a variation of $\mu$. It is noted that any wavelength-dependent distortion is likely to introduce a systematic error on the fiducial value of $\dmm$. This is because such distortion will produce a relative shift between the CO absorption features which mimics the effect of a non-zero $\dmm$ in \mbox{Equation \ref{eq:shift}}. This phenomenon is limited by fitting bands that have different $K_{i}$ values at similar wavelengths, as for the A-X($1-0$) and the d-X($5-0$) bands. However, only the perturbing d-X($5-0$) band matches this condition, limiting the effectiveness in breaking the degeneracy.

In recent years, UVES has been found to suffer from `intra-order distortions' -- wavelength-dependent velocity shifts whose pattern repeats across echelle orders, i.e. at scales of \mbox{$\sim$ 50-100 \AA} \citep{Whitmore2010,Whitmore2015}. \cite{Rahmani2013} also found that UVES suffers from wavelength calibration distortions on longer scales. A detailed investigation showed that such long-range wavelength distortions, on scales of \mbox{$\sim$ 1000-3000 \AA}, are ubiquitous across the entire history of UVES \citep{Whitmore2015}. It is commonly accepted that such distortions are due to different paths of the light-beam from the quasar and from the ThAr calibration lamp in the spectrograph and to the different illumination of the slit between the science object and the calibration lamp exposures (though the evidence for this remains unclear at present).

\subsubsection{Long-range wavelength distortions}
\label{subsubsec:longrange}
\cite{Molaro2008} first proposed the technique, now often referred to as `supercalibration', to correct the spectrum for such distortions. The technique, which was later improved and refined by \cite{Whitmore2015}, consists in a comparison between a UVES and a reference spectrum with a much more accurate frequency scale. The reference spectrum used in this work is the solar spectrum\footnote{\url{http://kurucz.harvard.edu/sun/irradiance2005/irradthu.dat}} taken with a Fourier Transform Spectrometer (FTS) and reported by \cite{Chance2010}. Typical targets for the supercalibration technique are asteroids and `solar-twin' stars, since the former reflect the spectrum of the Sun and the latter show a spectrum which is almost identical to that of the Sun \citep{Melendez2009,Datson2014}.

\begin{figure}
    \centering
    \includegraphics[width=1.\columnwidth]{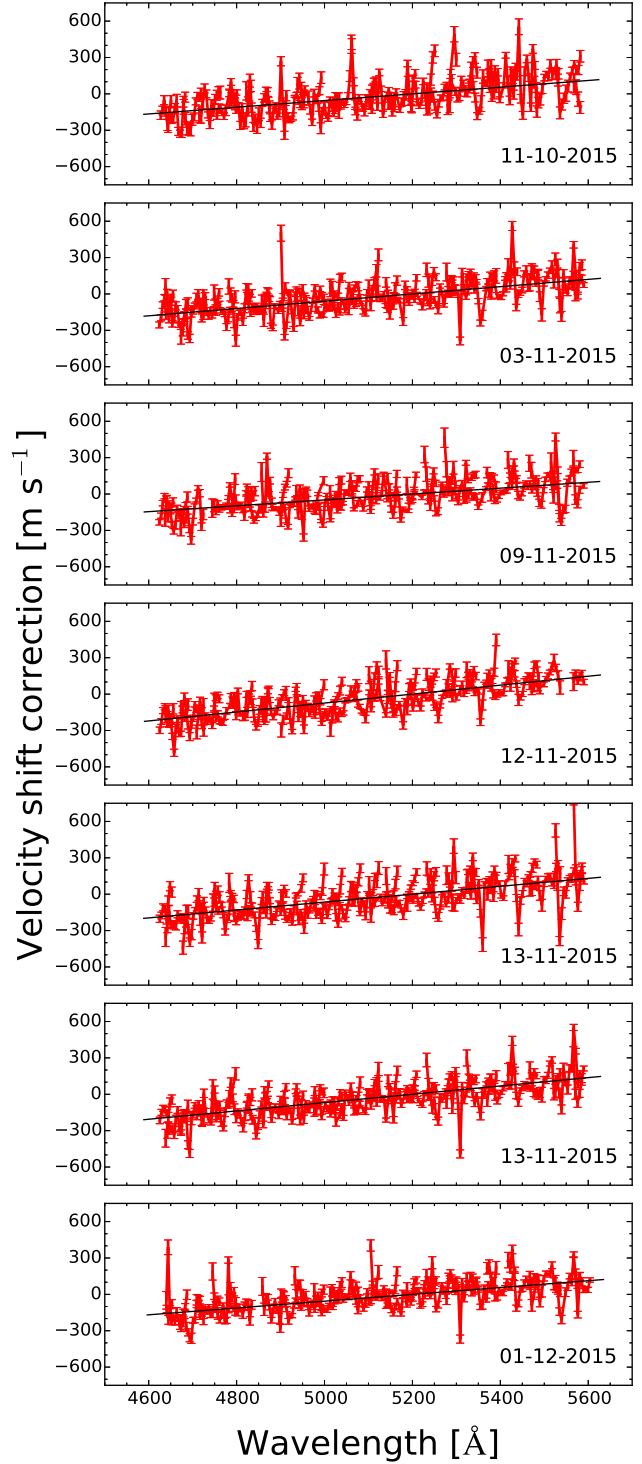}
    \caption{Map of the UVES long-range wavelength scale distortions in the solar twin HD001835 exposures taken in 2015, displaying a one-to-one correspondence with the values listed in \mbox{Table \ref{tab:dist}}, i.e. the top panel corresponds to the first row of the table. All the exposures were recorded immediately after their corresponding \jo\ science exposures. In each panel, the long-range distortions relative to a single quasar exposure are shown for the lower CCD in the red arm of UVES. The velocity shifts were measured on \mbox{$\sim 10$ echelle} orders in each exposure. The fitted slopes show the velocity shift needed to correct for the long-range distortions. Since any constant velocity offset is not relevant for the long-range distortions analysis, the map was shifted to a zero velocity at $\lambda = $ \mbox{5300 \AA}.}
    \label{fig:supercali}
\end{figure}

The \jo\ exposures taken in 2015 were distortion-corrected using dedicated solar twin supercalibrations taken immediately after the quasar exposure and its attached ThAr calibration exposure. Each quasar exposure was distortion-corrected using its dedicated supercalibration, following the same procedure as \cite{Bagdonaite2014} and \cite{Dapra2015,Dapra2016}. Briefly, the long-range velocity distortions are characterised by a single slope as a function of wavelength in each of the blue and red arms of UVES. The supercalibration velocity measurements are shown in \mbox{Fig. \ref{fig:supercali}} (for the most relevant red arm) and the measured values of the distortion slopes are listed in \mbox{Table \ref{tab:dist}} for the red and the blue arms.
\begin{table}
    \centering
    \caption{Details of the solar twin HD001835 supercalibration exposures taken in 2015. Each exposure was taken with the same settings of the relative science quasar exposure. The uncertainty on the slopes is \mbox{$\sim 30$ \ms} per 1000 \AA. Note that the CO bands are detected only in the red arm of UVES. The distortion slopes for the blue arm are included for completeness.}
    \label{tab:dist}
    \begin{tabular}{ccc}
    \hline
    Date & \multicolumn{2}{c}{Distortion slope [\ms\ per 1000 \AA]} \\
     & Blue arm & Red arm \\
    \hline
    \hline
    11-10-2015 & 350 & 280 \\
    03-11-2015 & 420 & 300 \\
    09-11-2015 & 380 & 240 \\
    12-11-2015 & 570 & 370 \\
    13-11-2015 & 440 & 330 \\
    13-11-2015 & 510 & 340 \\
    01-12-2015 & 590 & 280 \\
    \hline
    \end{tabular}
\end{table}

Since the exposures from 2014 were not recorded with dedicated supercalibrations, an attempt to distortion correct them was made following the same approach as \cite{Bagdonaite2014} and \cite{Dapra2017}. The ESO archive was inspected looking for asteroids and solar twin exposures taken within \mbox{$\sim$ 1 week} of the quasar exposures. More than 300 observations of the solar twin star HD217014 were recorded in August 2014 under the program 093.C-0929(D) (PI Martins). However, those exposures were recorded using only the red arm of UVES, with the lower red CCD centred at \mbox{580 nm}, a slit width of \mbox{0.3 arcsec} and a $1 \times 1$ CCD binning. In view of the narrower slit width used in the solar twin exposures, which may affect the path of the light from the solar twin with respect to the quasar exposures, the wavelength distortion corrections derived from the solar twin spectra may not be applied reliably to address the distortions in the CO spectrum. Nevertheless, an attempt was made to use the solar twin exposures for an estimate of the long-range wavelength distortions during August 2014. The comparison between the HD217014 spectrum and the reference solar spectrum returned distortion slope values in the range \mbox{90-130 \ms} per \mbox{1000 \AA}, with an average distortion slope of \mbox{$\sim 100$ \ms} per \mbox{1000 \AA}. The spectrum created by counter-distorting the exposures taken in 2014 with the average distortion slope values returned an updated fiducial value of \mbox{$\dmm = (1.8 \pm 2.2_{\stat}) \times 10^{-5}$}. The difference between the minimum and the maximum value for the distortion slope was interpreted as the uncertainty on the adopted value of the slope. This translates into a systematic uncertainty on $\dmm$ of \mbox{$\sim 0.2 \times 10^{-5}$}, which was added to the systematic error budget.

\subsubsection{Intra-order distortions}
\label{subsec:intra}
\cite{Whitmore2010} found that the wavelength scale of the ThAr calibration suffers from velocity shifts within each echelle order. These shifts have a magnitude of several hundreds of \ms\ and may affect the fiducial value of $\dmm$ by introducing a relative shift among the CO transitions. However, since the position of the CO transitions along their respective echelle orders is independent of the repeated pattern of intra-order distortions across the orders, the velocity shift imparted to a given CO transition can be considered as randomly distributed. 
The intra-order distortions introduce in each exposure a velocity shift which is translated into a systematic error on $\dmm$ by:
\begin{equation}
    \delta(\frac{\Delta\mu}{\mu}) = \frac{(\Delta v/c)}{\sqrt{N} \Delta K_{i}},
    \label{eq:intra}
\end{equation}
where $\Delta v$ is the mean magnitude of the intra-order distortions, $\Delta K_{i} = 0.06$ is the spread in the sensitivity coefficients, and \emph{N} is the number of the CO transitions. Because of the self-blending of the CO branches, the number of CO transitions that are effectively contributing to the signal is lower than the total number of CO transitions considered to build the band contour. By analyzing CO absorption from the same band systems in a different absorber, \cite{Dapra2016} found that only \mbox{$\sim 20\%$} of the total number of CO lines effectively contribute to the signal. As a consequence, a value of $N = 25$ was adopted for the number of transitions effectively contributing to the signal.

The dedicated supercalibration exposures taken in 2015 returned a mean amplitude of the intra-order distortions of \mbox{$\Delta v = 67.1$ \ms}, while the solar twin exposures from 2014 delivered a mean amplitude of \mbox{$\Delta v = 52.4$ \ms}. These values translate into $\dmm$ uncertainties of $\delta_{2015} = 0.34 \times 10^{-5}$ and $\delta_{2014} = 0.27 \times 10^{-5}$. The systematic uncertainty due to the intra-order distortions was derived by combining these two values using a weighted average, using the S/N of the two sub-spectra as weights. This procedure delivered an uncertainty on $\dmm$ of \mbox{$\sim 0.3 \times 10^{-5}$}, which was included in the systematic uncertainty budget.

\subsection{Spectral redispersion}
\label{subsec:redisp}
A potential source of systematic uncertainties is the spectral redispersion. While building the final 1D spectrum, the different exposures were redispersed onto a common wavelength grid. This procedure implies a rebinning of the spectra, which can cause flux correlations between adjacent pixels, and the (arbitrary) choice of the grid can slightly distort the line profiles, affecting the value of $\dmm$.

To estimate the impact of the spectral redispersion on the fiducial value presented in this work, the exposures were recombined using ten different wavelength grids in the range 1.9-2.1 \kms\ per pixel. Subsequently, a value for $\dmm$ was returned by each spectrum and was compared with the fiducial value. The average deviation from the fiducial valued was $0.1 \times 10^{-5}$, which was added to the systematic error budget.

\subsection{Total systematic uncertainty}
\label{subsec:total}
The total systematic uncertainty in the fiducial value of $\dmm$ was calculated by adding in quadrature all the contributions to the systematic error. The updated fiducial $\dmm$ value therefore becomes \mbox{$\dmm = (1.8 \pm 2.2_{\stat} \pm 0.4_{\syst}) \times 10^{-5}$}. This value is delivered by the analysis of 9 CO bands in the spectrum of \jo, which was corrected for the long-range wavelength distortions.

\section{Conclusion}
\label{sec:conclusion}
In this work, the analysis of CO absorption in the system at \mbox{$\zab = 2.52$} in the line-of-sight towards quasar \jo\ was presented, in order to constrain the temporal variation of the proton-to-electron mass ratio. CO was found in 10 different bands, although one is completely overlapped by intervening metal absorption and thus discarded, covering the wavelengths \mbox{4660-5448 \AA}.  A CO temperature of \mbox{$\tco=11.4$ K} was derived and, starting from the updated molecular database reported by \cite{Dapra2016}, effective band contours were created and fitted against the quasar spectrum using the comprehensive fitting technique. This approach allowed the simultaneous fitting of all the CO vibrational bands using four free parameters only, namely the total column density $\nco$, the absorption redshift $\zab$, the Doppler width $\bco$, and the relative variation of the proton-to-electron mass ratio $\dmm$. The absorption model returned $\dmm = (1.8 \pm 2.2_{\stat} \pm 0.4_{\syst}) \times 10^{-5}$. This value agrees well with the value of \mbox{$\dmm = (0.7 \pm 1.6_{\stat} \pm 0.5_{\syst}) \times 10^{-5}$} derived by \cite{Dapra2016} from the spectrum of SDSS~J123714.60+064759.5 in showing no variation of $\mu$ over a look-back time of \mbox{$\sim 11.2$ Gyrs}.

A way to improve this $\dmm$ value is to include the CO absorption in a combined analysis with different molecules, which are sensitive to a varying-$\mu$. Molecular hydrogen is the main candidate for such a combined analysis, since it is assumed to be cospatial with CO and it is often observed in $> 50$ transitions at $\zab > 2$. \cite{Dapra2016} reported the first combined analysis of CO and \htwo\ in the absorbing system towards quasar SDSS~J123714.60+064759.5. They found that the constraint on $\dmm$ returned by the combined analysis of CO and \htwo\ is $\sim 38\%$ more stringent than that derived from CO absorption only. Molecular hydrogen absorption in \jo\ was reported by \cite{Noterdaeme2017}, who detected it using an XShooter spectrum with a higher S/N and a lower resolution than UVES. However, the faintness of the background quasar results in a very low \mbox{$\textrm{S/N} < 2$} at the \htwo\ absorption wavelengths in the UVES spectrum (\mbox{$\sim 3600$ \AA}). Moreover, molecular hydrogen has a high column density \citep[$N(\textrm{H}_{2}) \sim 10^{20.43} \textrm{cm}^{-2}$,][]{Noterdaeme2017}, which results in strongly saturated transitions. Because of these two effects, \htwo\ absorption is unlikely to add any valuable signal to the analysis of $\dmm$ in this system.

\section*{Acknowledgments}
The authors thank the Netherlands Foundation for Fundamental Research of Matter (FOM) for financial support. MTM thanks the Australian Research Council for \textsl{Discovery Project} grant DP110100866 which supported this work. PN gratefully acknowledges support from the Indo-French Centre for the Promotion of Advanced research under grant 5504-2. WU thanks the European Research Council for an ERC-Advanced grant (No 670168). The work is based on observations with the ESO Very Large Telescope at Paranal (Chile).

\bsp
\label{lastpage}

\end{document}